\documentclass[12pt]{iopart}

\usepackage{graphicx}
\usepackage{mathrsfs}
\usepackage{amssymb}

\newcommand{\bra}[1]{\ensuremath{\left\langle#1\right|}}
\newcommand{\ket}[1]{\ensuremath{\left|#1\right\rangle}}

\begin{document}
\title[]{Counter-propagating spontaneous four wave mixing:  photon-pair factorability and ultra-narrowband single photons}
\author{Jorge Monroy-Ruz$^1$, Karina Garay-Palmett$^2$\footnote{Corresponding 
author: kgaray@cicese.mx}, and Alfred B. U'Ren$^1$}

\address{$^1$ Instituto de Ciencias Nucleares, Universidad Nacional Aut\' onoma de M\' exico,\\ Apartado Postal 70-543, 04510 DF, M\' exico}
\address{$^2$ Departamento de \' Optica, Centro de Investigaci\' on Cient\' ifica y de Educaci\' on Superior de Ensenada,
 Apartado Postal 360 Ensenada, BC 22860, M\' exico}
\ead{kgaray@cicese.mx}

\begin{abstract}
We introduce a new kind of spontaneous four wave mixing process for the generation of photon pairs, in which the four waves involved counter-propagate in a guided-wave $\chi^{(3)}$ medium; we refer to this process as counter-propagating spontaneous four wave mixing (CP-SFWM).    We show that for the simplest CP-SFWM source, in which all waves propagate in the same polarization and transverse mode and in which self- and cross-phase modulation effects are negligible, phasematching is attained \emph{automatically} regardless of dispersion in the fiber or waveguide.   Furthermore, we show that in two distinct versions of this source (both pumps pulsed, or one pump pulsed and the remaining one monochromatic), the two-photon state is \emph{automatically} factorable provided that the length of the nonlinear medium exceeds a certain threshold, easily achievable in practice since this threshold length tends to be in the range of mm to cm.  We also show that if one of the pumps approaches the monochromatic limit, and for a sufficient nonlinear medium length, the bandwidth of one of the two photons in a given pair may be reduced to the level of MHz,  compatible with electronic transitions for the implementation of atom-photon interfaces, without the use of optical cavities.  
\end{abstract}

\maketitle

\tableofcontents

\section{\label{sec.Intro}Introduction}

%
%

Photon pairs produced by spontaneous parametric processes have enabled many important advances in quantum-enhanced technologies such as quantum metrology \cite{Giovannetti11}, quantum communications \cite{Gisin07} and quantum computation \cite{Kok07}.  The processes of spontaneous parametric downconversion (SPDC) based on second-order non-linearities \cite{Burnham84} and of spontaneous four wave mixing (SFWM) based on third-order non-linearities \cite{Fiorentino02} are well-established as  sources of photon pairs.  The SFWM process, implemented in optical fibers \cite{Fulconis07, Cohen09, CruzDelgado14}, has gained prominence as a viable alternative to  SPDC with a number of distinct advantages including the elimination of losses associated with coupling of photon pairs into optical fibers, a greater scope for photon-pair engineering \cite{Garay07}, as well as the possibility of an essentially unlimited interaction length in long optical fibers.  
  
The development of photon-pair sources based on guided-wave non-linear optical media (fibers or waveguides) with tailored spatio-temporal properties is an ongoing field of research. On the one hand, it is well known that in order to herald a quantum-mechanically pure single photon from a photon-pair, it is essential that the two-photon quantum state be free from entanglement in all photonic degrees of freedom \cite{URen05}. While an appropriate combination of spectral and spatial filtering can render a two-photon state factorable, scalability to higher dimensions for protocols requiring multiple pure heralded single photons necessitates in practice photon-pair engineering so that filtering may be precluded.   On the other hand, while photon-atom interfaces require single photons with both frequency and bandwidth matched to those of the atomic transition in question, photon-pair sources based on both SPDC and SFWM tend to be characterized by a bandwidth which is orders of magnitude larger than that of atomic transitions.    In order to remedy this, one possibility is to resort to cavity-enhanced processes in which the nonlinear medium is placed inside a high-finesse optical cavity resulting in the emission of photon pairs in the (narrow) spectral modes supported by the cavity \cite{Ou99, Kuklewicz06, bao08, Jeronimo10, Fekete13, Garay13}.  

In the spontaneous four wave mixing (SFWM) process, two pump photons are annihilated in a guided-wave $\chi^{(3)}$ medium, such as a fiber or waveguide, leading to the generation of signal and idler photon pairs in such a manner that energy and momentum are conserved.   In all SFWM sources demonstrated to date, the four waves  involved (the two pumps, the signal, and the idler) propagate in the same direction along the fiber or waveguide.   In this paper we introduce a new kind of SFWM process,  to the best of our knowledge not studied previously, in which the two pump waves are launched from opposite ends so that they counter-propagate in the non-linear medium.  We refer to such a process as counter-propagating spontaneous four wave mixing, or CP-SFWM.  In this process,  one of the daugther photons, which we call signal,  is emitted so that it backpropagates with respect to pump 1, while the conjugate idler photon backpropagates with respect to pump 2.  Note that $\chi^{(2)}$-based processes have been studied in which the two generated photons counter-propagate leading to interesting spatio-temporal engineering possibilities  \cite{Walton03, torres04, tsang2005, christ2009, caillet09, boucher2014, gatti2015, corti2016}.  Also note that classical implementations of four wave mixing (stimulated process) with counter-propagating fields have been previously proposed and demonstrated \cite{yariv1977, jensen1978, bian2004, yan2011}.

As we discuss below, the CP-SFWM process leads to some unique properties that distinguishes it from standard SFWM. First, regardless of the specific dispersion properties of the non-linear medium, phasematching is automatically attained for all conceivable single-mode fibers (or waveguides), as long as the four waves are characterized by the same dispersion relation, at generation frequencies ($\omega_s$ and $\omega_i$) that match those of the pumps ($\omega_1$ and $\omega_2$), according to $\omega_s=\omega_1$ and $\omega_i=\omega_2$.    This symmetry is broken in the presence of self- and/or cross-phase modulation effects or if the four waves involve different polarizations and/or transverse modes leading to slight offsets from $\omega_s=\omega_1$ and $\omega_i=\omega_2$, thus facilitating experimental discrimination of the CP-SFWM photons from scattered pump photons.   
The automatic phasematching represents a considerable advantage as for any given optical fiber it becomes possible to freely choose the pump frequencies and thus also directly determine the generation frequencies, according to particular needs.  Second, as we show in detail below, unlike the case of standard SFWM for which factorability can be accomplished under highly restrictive group velocity matching conditions, involving certain specific combinations of frequencies, in the case of CP-SFWM factorability is accomplished for any phasematched source design, as long as the nonlinear medium length exceeds a certain threshold.   Automatic phasematching \emph{and} automatic factorability indeed become a powerful combination in photon-pair source design.   Third, we show below that when making one of the two pumps nearly monochromatic and if the nonlinear medium length exceeds a certain threshold, CP-SFWM also permits the generation of photon pairs for which one of the two photons can be characterized by an ultranarrow bandwidth, without resorting to the use of optical cavities.

\section{\label{Theory}Theory of counterpropagating SFWM }

\begin{figure}[t]
\centering
\includegraphics[width=10cm]{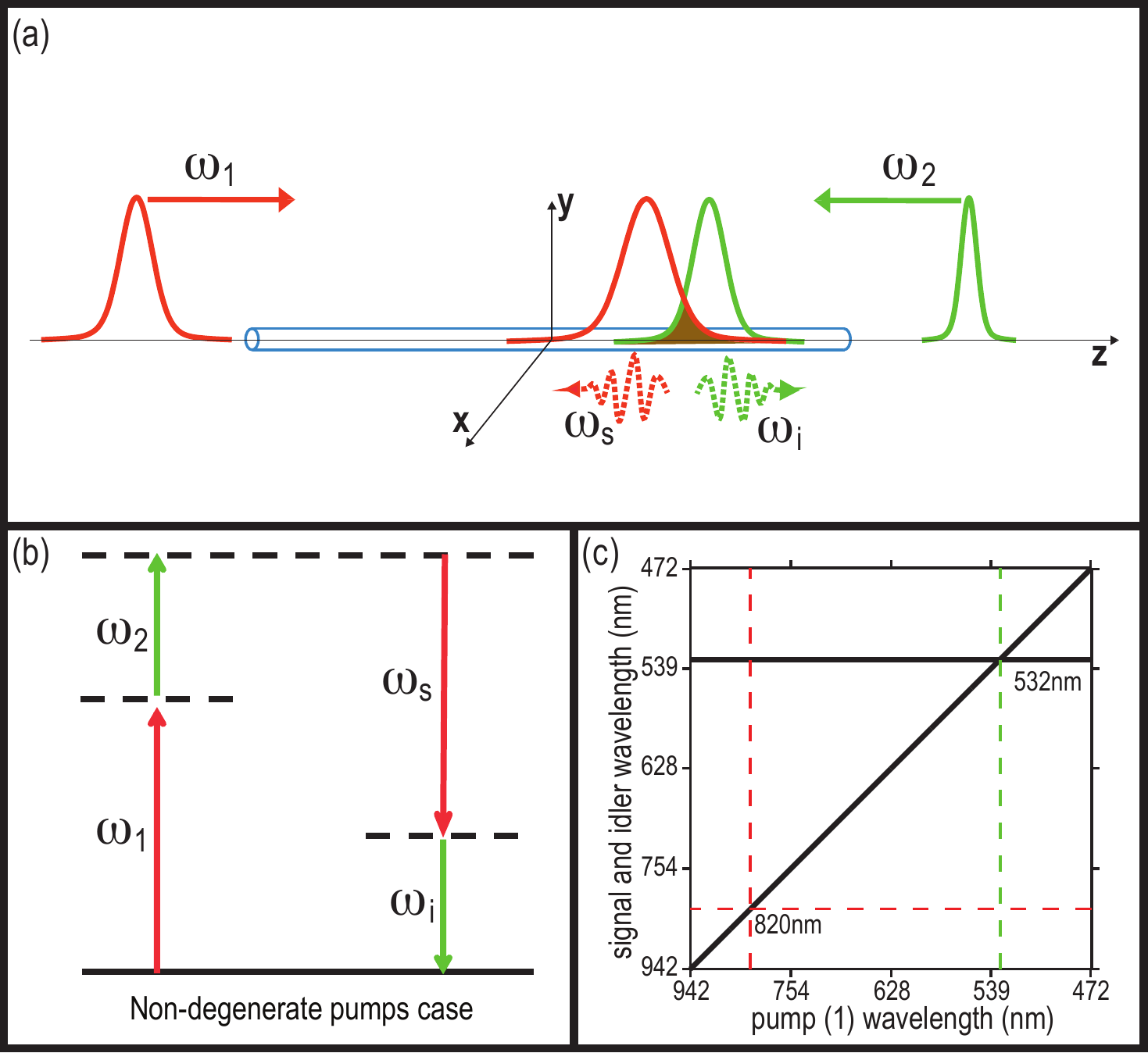}
\caption{(a) Schematic of the CP-SFWM process.  The gaussian-shaped pumps are represented in solid red (pump 1) and solid green (pump 2), while the generated CP-SFWM photons are indicated in dashed red (signal) and dashed green (idler);  arrowheads indicate the directions of propagation for the four waves. (b) Energy level diagram of the process. (c) Phasematching diagram. The solid black lines represent the signal and idler frequencies that fulfil phasematching, as a function of $\omega_{1}$ and for $\omega_2$ fixed at the particular value $\omega_2 = (2\pi c)/0.532 \mu m$.}
\label{FigEsquema}
\end{figure}

While all conclusions reached in this paper could apply to both waveguide and fiber sources, henceforth we refer to the nonlinear medium as `fiber' with the understanding that it could equally refer to a waveguide.   Photon-pair generation experiments based on the process of spontaneous four wave mixing demonstrated to date involve four waves, i.e. pump 1, pump 2, signal, and idler, which propagate along the fiber in the same direction, see for example \cite{Fiorentino02,Fulconis07,Cohen09,CruzDelgado14}. Here, we propose a SFWM scheme in which the pump fields counter-propagate, i.e. they are launched into the fiber from opposite ends. In such a SFWM interaction, a photon from the pump at frequency $\omega_1$ and travelling in the forward direction, together with and a photon from the pump at frequency $\omega_2$  travelling in the backward direction, are annihilated giving rise to the emission of a counterpropagating photon pair, which is a consequence of energy and momentum conservation constraints. The generated pair is comprised of a backward-propagating signal photon at frequency $\omega_s$ and a forward-propagating idler photon at frequency $\omega_i$. The described interaction is illustrated in figure~\ref{FigEsquema}.  

\subsection{\label{sec.State}The two-photon state}

In this section we describe the two-photon state for the CP-SFWM process in a $\chi^{(3)}$ medium. We will initially write down expressions for the two-photon state which permit each of the four waves to propagate in different transverse and polarization modes, where $k_1(\omega)$, $k_2(\omega)$, $k_s(\omega)$, and $k_i(\omega)$ represent the frequency-dependent wavenumbers for each of the four waves: pump 1(1), pump 2(2), signal(s), and idler(i).  Later  in the paper we will concentrate our discussion on the case where all four waves are co-polarized and involve the same transverse mode so that $k_1(\omega)=k_2(\omega)=k_s(\omega)=k_i(\omega)\equiv k(\omega)$.  Throughout this paper, while the pump $1$ and the idler waves are forward-propagating, the pump $2$ and signal waves are backward propagating; we adopt a sign convention for which all wavenumbers are positive, with explicit signs appearing in accordance to the direction of propagation.

We start from the interaction Hamiltonian governing SFWM processes,  given by

\begin{equation}
\label{Hamilt}\hat{H}(t)=\frac{3}{4}\epsilon_o\chi^{(3)} \!\int\!\! d^3 \textbf{r} \hat{E}_{1}^{(+)}(\textbf{r},t)\hat{E}_{2}^{(+)}(\textbf{r},t)\hat{E}_{s}^{(-)}(\textbf{r},t)\hat{E}_{i}^{(-)}(\textbf{r},t),
\end{equation}

\noindent where the integration is carried out over the portion of the nonlinear medium for which the pump fields are temporally and spatially overlapped, $\chi^{(3)}$ is the third-order nonlinear susceptibility,  and $\epsilon_o$ is the vacuum electrical permittivity.  In Eq. (\ref{Hamilt}), the subscripts $^{(+)}/^{(-)}$ refer to the positive frequency / negative frequency parts  of the electric field operators.  In our analysis, we assume that the two pumps can be well-described by classical fields, i.e. no longer operators, of the form

\begin{equation}
\label{pump}\hat{E}_{\nu}^{(+)}(\textbf{r},t) \rightarrow A_{\nu}f_{\nu}\left(x,y\right)\int\!\!d\omega\alpha_{\nu_\pm}(\omega)\,\mbox{exp}\left[-i\left(\omega t\mp k(\omega)z\right)\right],
\end{equation}

\noindent with $\nu=1,2$ for the two pumps and  $A_{\nu}$ represents the field amplitude.  $\alpha_{\nu_\pm}(\omega)$  is the spectral envelope (the meaning of the signs $\pm$  is defined below), and $f_{\nu}\left(x,y\right)$ is the transverse spatial field distribution, which is normalized so that $\int\!\!\int|f_{\nu}(x,y)|^2\,dxdy=1$ and is approximated to be frequency-independent within the pump bandwidth.

The quantized signal and idler fields are expressed as

\begin{equation}
\label{Ecuant}\hat{E}_{\mu}^{(+)}(\textbf{r},t)=i\sqrt{\delta k}f_{\mu}(x,y)\nonumber\sum_{k}\mbox{exp}\left[-i(\omega t\mp k(\omega)z)\right]\ell(\omega) \hat{a}_{\mu \pm}(k),
\end{equation}

\noindent with $\mu=s,i$ and $\delta k=2\pi/L_Q$ the mode spacing, written in terms of the quantization length $L_Q$. Function $\ell(\omega)$ is given as follows

\begin{equation}
\label{ldek}\ell(\omega)=\sqrt{\frac{\hbar \omega}{\pi\epsilon_o n^2(\omega)}},
\end{equation}

\noindent in terms of the (linear) refractive index of the nonlinear medium $n(\omega)$ and of Planck's constant $\hbar$. In Eq.~(\ref{Ecuant}), $\hat{a}_{\mu \pm }(k)$ is the annihilation operator (the meaning of the signs $\pm$  is defined below), and $f_{\mu}(x,y)$ represents the transverse spatial distribution of the field, which is also normalized as the corresponding pump functions, and is assumed to be frequency-independent within the bandwidth of signal and idler modes.

Note that in equations (\ref{pump}) and (\ref{Ecuant}) the  $-/+$ signs, in front of the propagation constant $k(\omega)$ and the corresponding subscripts $+/-$ in the annihilation operators and the pump spectral envelopes, indicate optical fields propagating along the fiber in the forward/backward directions.

Following a standard perturbative approach \cite{MandelWolf} and our treatment in reference \cite{Garay08}, it can be shown that the two-photon state produced by CP-SFWM  can be written as $\ket{\Psi}=\ket{0}_s\ket{0}_i+\eta\ket{\Psi}_2$, in terms of the two-photon component

\begin{equation}\label{2photonstate}
\label{state}\ket{\Psi}_2=\sum_{k_{s}}\sum_{k_{i}}\ell(k_{s})\ell(k_{i})F(k_{s},k_{i})\hat{a}^{\dagger}_{s-}(k_{s})\hat{a}^{\dagger}_{i+}(k_{i})\ket{0}_s\ket{0}_i,
\end{equation}

\noindent and the constant $\eta$, which is related to the conversion efficiency and is given by

\begin{equation}
\label{eta}\eta=i(2\pi)\delta k\frac{3\epsilon_o \chi^{(3)}}{4\hbar}A_1A_2Lf_{eff},
\end{equation}

\noindent where $L$ is the fiber length and $f_{eff}$ is the spatial overlap integral between the four fields given by

\begin{equation}
\label{overlap} f_{eff} =\int\! dx\!\int\! dyf_1(x,y)f_2(x,y)f^{\ast}_s(x,y)f^{\ast}_i(x,y).
\end{equation}

In Eq.~(\ref{state}) $k_{s}\equiv k_s( \omega_{s} )$ is the propagation constant for the backward-propagating signal mode, and $k_{i}\equiv k_i( \omega_{i} )$ is the propagation constant for the forward-propagating idler mode; $\hat{a}^{\dagger}_{s-}(k)$ represents the creation operator for the backward-propagating signal mode, while $\hat{a}^{\dagger}_{i+}(k)$ represents the creation operator for the forward-propagating idler mode. $F(k_{s},k_{i})$ is the joint amplitude function, which can be written in terms of frequencies rather than wave numbers, in which case it is referred to as the joint spectral amplitude (JSA), and is expressed as $F(\omega_{s},\omega_{i})$.

In our analysis we first consider a source configuration in which both pumps are pulsed.  In this case, the joint spectral amplitude function  $F_{\it{P}}(\omega_s,\omega_i)$ can be shown to be given by

\begin{equation}
\label{JSAp}F_{\it{P}}(\omega_{s},\omega_{i})=\int d\omega\alpha_{1+}(\omega)\alpha_{2-}(\omega_{s}+\omega_{i}-\omega)\mbox{sinc}\left[\frac{L}{2}\Delta k\right]e^{i\frac{L}{2} \kappa}e^{i\omega \tau},
\end{equation}

\noindent where $\alpha_{1+}(\omega)$ represents the pump spectral envelope for the forward-propagating pump,  $\alpha_{2-}(\omega)$ represents the pump spectral envelope for the backward-propagating pump, and  $\tau$ represents the time of arrival \emph{difference} between the two pump pulses at their respective ends of the fiber.  Note that $\tau$ can be controlled externally with a relative delay between the two pumps; in particular, $\tau=0$ implies that the pump pulses corresponding to pumps $1$ and $2$ arrive at the same time at the two ends of the fiber.    Eq. (\ref{JSAp}) is expressed in terms of the phase mismatch function $\Delta k\equiv\Delta k(\omega,\omega_s,\omega_i)$, and the function $\kappa\equiv \kappa(\omega,\omega_s,\omega_i)$ defined as

\begin{eqnarray}
\label{PMp}
\Delta k= k_1(\omega)- k_2(\omega_{s}+\omega_{i}-\omega)- k_s(\omega_{s})+k_i(\omega_{i})+\phi_{NL},\\
\kappa= k_1(\omega)+ k_2(\omega_{s}+\omega_{i}-\omega)+k_s(\omega_{s})+k_i(\omega_{i}),
\end{eqnarray}

\noindent  where $\phi_{NL}$ is a nonlinear phase shift derived from self-phase and cross-phase modulation (see below for further discussion and for expressions).  Note that the energy conservation constraint is already included in the resulting joint amplitude.

Let us now consider a pumps configuration defined as the limit where the backward-propagating pump wave becomes monochromatic at frequency $\omega_{cw}$, in which case the corresponding electric field can be expressed as $E_{cw}^{(+)}(\textbf{r},t)=af_2\left(x,y\right)\mbox{exp}\left[-i\left(\omega_{cw} t+k(\omega_{cw})z\right)\right]$ with $a$ the field amplitude, while the forward-propagating pump remains broadband; we refer to this as the mixed pumps configuration.   In this case, the JSA function becomes 

\begin{equation}
\label{JSAm}F_{\it{M}}(\omega_{s},\omega_{i})=\alpha_{+}(\omega_{s}+\omega_{i}-\omega_{cw})\mbox{sinc}\left[\frac{L}{2}\Delta k_{\it{M}}\right]e^{i\frac{L}{2}\kappa_{\it{M}}},
\end{equation}

\noindent  where  $\Delta k_{\it{M}}\equiv\Delta k_M(\omega_s,\omega_i)$ and $ \kappa_{\it{M}}\equiv\kappa_{\it{M}}(\omega_s,\omega_i)$ are defined as

\begin{eqnarray}
\label{PMm}
\Delta k_{\it{M}}= k_1(\omega_{s}+\omega_{i}-\omega_{cw})- k_2(\omega_{cw})- k_s(\omega_{s})+k_i(\omega_{i})+\phi_{NL},\\
 \kappa_{\it{M}}= k_1(\omega_{s}+\omega_{i}-\omega_{cw})+ k_2(\omega_{cw})+k_s(\omega_{s})+k_i(\omega_{i}).
\end{eqnarray}

The nonlinear phase shift $\Phi_{NL}$, appearing in Eq. (\ref{PMp}) (for the pulsed pumps case), can be been shown to be given as follows~\cite{Garay07, Garay2011r}

\begin{eqnarray}\label{phiNL}
\Phi_{NL}=(\gamma_1-2\gamma_{21}-2\gamma_{s1}+2\gamma_{i1})
P_1 -(\gamma_2-2\gamma_{12}+2\gamma_{s2}-2\gamma_{i2}) P_2,
\end{eqnarray}

\noindent where $P_1$ and $P_2$ represent the peak powers for pumps $1$ and $2$ (related, for pumps with Gaussian spectra, as $P_\nu=p_\nu \sigma_\nu / [\sqrt{2 \pi} R]$ with the average pump powers $p_\nu$, where $R$ is the repetition frequency and $\sigma_\nu$ is the corresponding bandwidth).   The nonlinear phase shift $\Phi_{NL}$, appearing in Eq. (\ref{PMm}) (for the mixed pumps case) is given by Eq. (\ref{phiNL}), with the substitution $P_2 \rightarrow p_2$.   

The coefficients $\gamma_{1}$ and $\gamma_{2}$ result from self-phase
modulation (SPM)  of the two pumps, and are given, with $\nu=1,2$ by

\begin{equation}\label{gam1}
\gamma_{\nu}=\frac{3\chi^{(3)}\omega_{\nu}^0f^{\nu}_{eff}}{4\epsilon_0c^2n_{\nu}^2}.
\end{equation}

In Eq.~(\ref{gam1}), the refractive index $n_{\nu}\equiv n(\omega_{\nu}^0)$ and the spatial overlap integral $f_{eff}^{\nu}\equiv \int\!\int\!dxdy|f_{\nu}(x,y)|^4$ (where the integral
is carried out over the transverse dimensions of the fiber)
are defined in terms of the carrier frequency $\omega_{\nu}^0$ for
pump-mode $\nu$ \cite{Agrawal}.  

In contrast, coefficients $\gamma_{\mu\nu}$
($\nu=1,2$ and $\mu=1,2,s,i$) correspond to the cross-phase
modulation (CPM) contributions that result from the dependence of the
refractive index experienced by wave $\mu$,  with $\mu=1,2,s,i$, on
the pump intensities ($\nu=1,2$;  note that CPM of the signal/idler photons on the pumps as well as
SPM of the signal and idler waves are small effects which we neglect).   These coefficients
are given by

\begin{equation}\label{gam2}
\gamma_{\mu\nu}=\frac{3\chi^{(3)}\omega_{\mu}^0f^{\mu\nu}_{eff}}{4\epsilon_0c^2n_{\mu}n_{\nu}},
\end{equation}

\noindent where $n_{\mu,\nu}\equiv n(\omega_{\mu,\nu}^0)$ is
defined in terms of the central frequency $\omega_{\mu,\nu}^0$ for
each of the four participating fields, and $f_{eff}^{\mu\nu}\equiv
\int\!\int\!dxdy|f_{\mu}(x,y)|^2|f_{\nu}(x,y)|^2$ is the
two-mode spatial overlap integral  (note that
$f_{eff}^{\mu\nu}=f_{eff}^{\nu\mu}$).

\subsection{\label{sec.flux}Expressions for the emitted flux }

In designing two-photon sources, it is helpful to be able to estimate the source brightness in terms of all relevant experimental parameters. Expressions for the emitted flux for co-propagating and co-polarized SFWM in single-mode fibers have been reported by us previously \cite{Garay08}. 

The number of photon pairs generated per second, or source brightness,  is given by
 
 \begin{equation}\label{defbrightness}
\label{NumPhot} N = R\sum_k\bra{\psi_2}\hat{a}_{s-}^{\dagger}(k)\hat{a}_{s-}(k)\ket{\psi_2},
\end{equation}
 
\noindent where $\ket{\psi_2}$ is defined in equation (\ref{state}) and $R$ is the pump repetition rate (for the pulsed pump in the mixed pumps case, and assumed to be equal for both pumps in the pulsed pumps case); note that the brightness in  Eq. (\ref{defbrightness}) can likewise be expressed in terms of the idler annihilation operator.   From this equation we have derived expressions for the number of photon pairs generated per second for the two pump configurations described above, which are represented by $N_{\it{P}}$ and $N_{\it{M}}$, respectively. In the analysis we assume that the pulsed pump fields have a gaussian spectral envelope 

\begin{equation}
\label{envSpec}\alpha_\mu(\omega)=\frac{2^{1/4}}{\pi^{1/4}\sqrt\sigma_\mu}\,\exp
\left[-\frac{(\omega-\omega^0_\mu)^2}{\sigma_\mu^2}\right],
\end{equation}

\noindent where $\omega^0_\mu$ and $\sigma_\mu$ (with $\mu=1,2$) are the central pump frequency and the pump bandwidth for the two pumps, respectively. Note that the function $\alpha_\mu(\omega)$ has been normalized so that $\int\!d\omega\! \mid\alpha_\mu(\omega)\mid^2=1$.

Substituting the two-photon state, Eq. (\ref{2photonstate}), into Eq. (\ref{defbrightness}) while appropriately turning sums into integrals in the limit $\delta k \rightarrow 0$, it can be shown  that $N_{\it{P}}$ is given by

\begin{equation}
\label{NumPhot_P} N_{\it{P}}=\frac{2^5n_{1}n_{2}c^2L^2\gamma^2p_1p_2}{\pi^3\omega_{1}^0\omega_{2}^0\sigma_1\sigma_2R}\int\! d\omega_s\!\int\! d\omega_i\,h(\omega_s,\omega_i)\mid F_{\it{P}}(\omega_s,\omega_i)\mid^2,
\end{equation}

\noindent where $n_1\equiv n_1(\omega_1^0)$ ($n_2\equiv n_2(\omega_2^0)$), $c$ is the speed of light in the vacuum, $L$ is the fiber length, $p_{\nu}$ and $\sigma_{\nu}$ (with $\nu=1,2$) are the average power and bandwidth of the two pumps, respectively, $F_{\it{P}}(\omega_s,\omega_i)$ is the JSA given in the equation (\ref{JSAp}), and  $h(\omega_s,\omega_i)$ is a function defined as

\begin{equation}
\label{funh} h(\omega_s,\omega_i)=\frac{\omega_sk'_s(\omega_s)}{n_s^2(\omega_s)}\frac{\omega_ik'_i(\omega_i)}{n_i^2(\omega_i)},
\end{equation}

\noindent where $k'_{\mu}(\omega_{\mu})$ denotes the frequency derivate of the propagation constant $k_{\mu}(\omega_{\mu})$, and $\gamma$ is the SFWM nonlinear coefficient given by

\begin{equation}
\label{gamma} \gamma = \frac{3\chi^{(3)}\sqrt{\omega_{1}^0\omega_{2}^0}f_{eff}}{4\epsilon_0c^2n_{1}n_{2}},
\end{equation}

\noindent in terms of the third-order nonlinear susceptibility, $\chi^{(3)}$, the electric permittivity of free space, $\epsilon_0$, and the spatial overlap integral $f_{eff}$ defined in Eq. (\ref{overlap}). Note that in our analysis we have assumed that the transverse electric field distributions for the various fiber modes depend only weakly on the frequency, so that $\gamma$ has been regarded as a constant, and taken out of the integral, as in  Eq. (\ref{NumPhot_P}).

Similarly, it can be shown that for the mixed pumps case the number of photon pairs generated per second, $N_{\it{M}}$, is given by

\begin{equation}
\label{NumPhot_M} N_{\it{M}}=\frac{2^{11/2}n_{1}n_{2}c^2L^2\gamma^2p_1p_2}{\pi^{3/2}\omega_{1}^0\omega_{2}\sigma}\int\! d\omega_s\!\int\! d\omega_i\,h(\omega_s,\omega_i)\mid F_{\it{M}}(\omega_s,\omega_i)\mid^2,
\end{equation}

\noindent where $p_1$ and $\sigma$ represent the average power and bandwidth of the pulsed pump, respectively; $p_2$ is the power of the monochromatic pump wave; $F_{\it{M}}(\omega_s,\omega_i)$ is the JSA given in the equation (\ref{JSAm}); $h(\omega_s,\omega_i)$ is given by equation (\ref{funh}) and $\gamma$ is defined according to equation (\ref{gamma}).

\subsection{\label{sec.PM}Phasematching properties and SFWM-pump discrimination}

In order for the CP-SFWM process to exist, linear momentum must be conserved which is equivalent to a phasematching condition $\Delta k =0$ for the pulsed pumps case, or  $\Delta k_M=0$ for the mixed pumps case (note that since in general all four waves are polychromatic, these phasematching conditions are fulfilled exactly for specific frequencies regarded as ``central'' for each wave).   It is straightforward to verify from Eqs. (\ref{PMp}) and (\ref{PMm}) that if all four waves propagate in the same transverse and polarization mode, then phasematching is always attained, provided that the nonlinear term $\phi_{NL}$ is negligible, at frequencies satisfying the following relationships:   $\omega_1=\omega_{s}$ and $\omega_2=\omega_{i}$. This is a remarkable property of CP-SFWM: phasematching is fulfilled \emph{automatically}, for an arbitrary single-mode fiber, using frequency non-degenerate pumps centered at $\omega_1$ and $\omega_2$ so as to generate a backward-propagating signal photon with frequency $\omega_{s}=\omega_1$ paired with a forward-propagating idler photon with frequency $\omega_{i}=\omega_2$.  In other words, basic phasematching properties (i.e. the determination of emission SFWM frequencies as a function of pump frequencies), become decoupled from the fiber dispersion and are in fact identical for all conceivable single-mode fibers.  Note that a particular case of the scenario above is that for which the pumps are frequency-degenerate which in fact leads to all four waves being frequency degenerate.   Also, note that a possible source of noise in CP-SFWM is spontaneous Brillouin scattering of the pump fields, which would appear in the same frequencies and directions of propagation as the generated photons \cite{Agrawal}.  

In figure \ref{FigEsquema}(c) we have plotted the $\Delta k=0$ contour (solid black straight lines), i.e. the signal and idler frequencies which satisfy perfect phasematching, as a function of the pump frequency $\omega_1$, while $\omega_2$ remains fixed at $\omega_2=2\pi c/0.532\mu$m.  Note that this diagram is ``universal'', in the sense that it applies to all conceivable single-mode fibers.  While a fixed pump 2 frequency leads to an equally fixed idler frequency, since $\omega_i=\omega_2$, there is a linear dependence between the remaining two frequencies, as $\omega_s=\omega_1$.
Note also that the intersection of the two straight lines  corresponds to the degenerate pumps case, for which $\omega_1=\omega_2=2\pi c/0.532\mu$m.  In particular, in the figure \ref{FigEsquema}(c) the red vertical line corresponds to $\omega_1=2\pi c/0.820\mu$m, so that its intersection with the $\Delta k=0$ contour indicates that perfect phasematching occurs for $\omega_s=2\pi c/0.820\mu$m and $\omega_i=2\pi c/0.532\mu$m.   A different choice of fixed pump $2$ frequency would simply lead to a vertically-displaced horizontal tuning curve for the idler photon.  Let us emphasize that the automatic phasematching observed for CP-SFWM  is \emph{achromatic} in the sense that it is attained for any choice of $\omega_1$ and $\omega_2$ (leading to $\omega_s=\omega_1$ and $\omega_i=\omega_2)$, regardless of the specific underlying fiber dispersion. This opens a wealth of possibilities for the implementation of photon-pair sources in optical fibers.

Note that if the nonlinear term $\phi_{NL}$  is non-zero, the symmetry between each pair of counterpropagating pump and generated SFWM photon is broken; in principle, this could  be useful in order to slightly offset the generation frequencies from the pump frequencies so as to simplify the experimental discrimination of signal and idler photons from scattered pump photons.  Likewise, this symmetry can be broken with cross-polarized SFWM processes of the kind $xyxy$ or $xxyy$ in brirefringent fibers.    However, for  experimental conditions regarded as typical (conventional fibers and typical values of pump power and/or typical brifrefringence values)  the resulting offset tends to be insufficient in practice for the effective discrimiantion between SFWM photon pairs and pump photons.  

Another interesting possibility is for the four waves to propagate in different transverse modes (in a few-mode or multi-mode fiber), likewise leading to an offset of the generation frequencies from the pump frequencies.    Thus, let us discuss the case of CP-SFWM implemented in a few-mode optical fiber \cite{Garay16,Cruz2016} leading to an intermodal process; this means allowing some of the waves involved in the SFWM process to travel in higher-order transverse modes. As an example, if the forward propagating pump travels in the fundamental fiber mode and the signal photon travels in a certain higher-order mode $X$, while the backward-propagating pump mode travels in the same higher-order mode $X$ and the idler photon travels in the fundamental mode,  then perfect phasematching will occur for signal and idler frequencies that are shifted from those of the pumps.  Specifically, this will result in a signal photon with frequency $\omega_1+\delta$ and in an idler photon with frequency  $\omega_2-\delta$, with a frequency offset $\delta$ which depends on the dispersion relations of the two modes; note that while the frequency offsets are equal (opposite in sign), the resulting wavelength offsets will differ between signal and idler.   Importantly,  as the order of mode $X$ increases, $\delta$ also increases. In table \ref{Tablamodes} we summarize the emission wavelengths and the resulting wavelength offsets that result from intermodal CP-SFWM in a step-index fiber (with numerical aperture $NA=0.3$ and core radius $r=2\mu$m)  that supports three higher-order modes, for the case in which the pump wavelengths are $820$nm (forward propagating pump) and $532$nm (backward propagating pump).

\begin{table}[ht]
\begin{center}
\begin{tabular}{| c | c | c | c | c |   }
  \hline         
  Fiber mode $X$		& $\lambda_s$ (nm)	&  $\Delta \lambda_s$(nm) 	& $\lambda_i$ (nm) 	& $\Delta \lambda_i$(nm) \\  \hline
  $LP_{11}$   		& 816.1 			& -3.9 					& 533.7 			& 1.7 \\  \hline   
  $LP_{21}$   		& 811.1 			& -8.9 					& 535.8 			& 3.8  \\  \hline    
  $LP_{02}$   		& 809.7 			& -10.3 					& 536.4		        & 4.4  \\  \hline  
  \hline  
\end{tabular}
\caption{Emission wavelengths ($\lambda_s$ and $\lambda_i$) and wavelength offsets ($\Delta \lambda_s$ and $\Delta \lambda_i$) for intermodal CP-SFWM, for different choices of excited mode $X$, in a few-mode step-index fiber with numerical aperture $NA=0.3$ and core radius $r=2\mu$m.}\label{Tablamodes}
\end{center}
\end{table}

We point out that while we have verified that the conclusions reached in this paper about factorability and ultra-narrowband single photon generation are unaffected by the use of 
the intermodal CP-SFWM process described above for signal/idler-pumps discrimination, for simplicity in the rest of the paper we concentrate on a CP-SFWM process which utilizes a single transverse mode.

\subsection{\label{sec.analytical}Closed analytical expressions for the joint spectral amplitude and the emitted flux}

In this section we show that under certain approximations it becomes possible to derive analytical expressions, in closed form, for both the joint spectral amplitude and for the emitted flux. Specifically, these approximations involve: i) writing the propagation constant $k(\omega)$, for each of the four interacting fields, as a first-order Taylor expansion around the frequencies for which perfect phasematching is obtained, and ii) assuming that the function $h(\omega_s,\omega_i)$ (see eq. (\ref{funh})) varies slowly within the spectral range of interest, so that we can regard it as a constant when evaluating the integrals in equations (\ref{NumPhot_P}) and (\ref{NumPhot_M}) in the section \ref{sec.flux}.   Note that  these approximations are no longer valid for large spectral spreads of the signal and idler frequencies around the frequencies which yield perfect phasematching.


For the pulsed pumps case (assumed to be Gaussian in spectrum, see equation (\ref{envSpec})), under the approximations mentioned above, and  using the integral form of the $\mbox{sinc}$ function

\begin{equation}
\label{sincF} \mbox{sinc}(x)=\frac{1}{2}\int_{-1}^1d\xi e^{ix\xi},
\end{equation}

\noindent the integral in equation (\ref{JSAp}) can be carried out analytically resulting in the approximate expression for the joint spectral amplitude  $ f^{lin}_{\it{P}}(\nu_s,\nu_i)=\alpha_{\it{P}}(\nu_s,\nu_i)\phi_{\it{P}}(\nu_s,\nu_i)$, where $\alpha_{\it{P}}(\nu_s,\nu_i)$ is determined by the two pump waves and is given by 

\begin{equation}
\label{alpha_Pul} \alpha_{\it{P}}(\nu_s,\nu_i)=\mbox{exp}\left[-\frac{(\nu_s+\nu_i)^2}{\sigma_1^2+\sigma_2^2}\right],
\end{equation}

\noindent  while $\phi_{\it{P}}(\nu_s,\nu_i)$ is determined both by the pump waves and the properties of the fiber, and has the form

\begin{equation}
\label{phi_Pul}  
\phi_{\it{P}}(x)=\mbox{exp}\left[-B^2x^2\right]\left[\mbox{erf}\left(\frac{1+\Lambda}{4B}+iBx\right) +\mbox{erf}\left(\frac{1-\Lambda}{4B}-iBx\right)\right],
\end{equation}

\noindent given in terms of the variable $x$, and parameters $B$ and $\Lambda$, defined as 

\begin{eqnarray}
\label{x}  x=T_s\nu_s+T_i\nu_i,\\ \label{B} 
                 B=\frac{\sqrt{\sigma_1^2+\sigma_2^2}}{t_{12}\sigma_1\sigma_2}, \\
                 \Lambda=\frac{1}{t_{12}}(2\tau +\tau_{12}),
\end{eqnarray}

\noindent where $\mbox{erf(.)}$ denotes the error function, $\nu_{\mu} = \omega_{\mu}-\omega_{\mu}^0$ are detuning variables (${\mu}=s, i)$, and $T_s$, $T_i$,  $t_{12}$, and $\tau_{12}$ are defined in tables \ref{Tpar}; $\tau$ was defined in the context of Eq. (\ref{JSAp}).   Note that here $\omega_{\mu}^0$ (with $\mu=1,2,s,i$) represent the central frequencies of the four waves involved . The definitions provided in tables \ref{Tpar} correspond to temporal variables; $\tau_{ij}$ terms represent transit time \emph{differences} through the fiber 
between waves $i$ and $j$, while $t_{ij}$ terms represent transit time \emph{sums} through the fiber between waves $i$ and $j$. In conventional fibers, with a length of a few cm,   $t_{ij}$ is on the order of tenths of nanoseconds, while $\tau_{ij}$ is on the order of few picoseconds.

In the left-hand-side table we have shown the various temporal parameters which define the two-photon state in the general case where the four waves may involve different polarizations and transverse modes.  In the right-hand-side table, we have specialized this to the case for which all four waves involve the same polarization and the same transverse mode.

{\renewcommand{\arraystretch}{1.6}
\begin{table}[t]
\begin{center}
\begin{tabular}{| c | c | }
  \hline   
  $T_s=t_{2s}-\frac{\sigma_1^2}{\sigma_1^2+\sigma_2^2}t_{12}$     & $T_{i}=\tau_{2i}-\frac{\sigma_1^2}{\sigma_1^2+\sigma_2^2}t_{12}$   \\  \hline
  $t_{12}  = L(k^{\prime}_1+k^{\prime}_2)$    &  $\tau_{12}= L(k^{\prime}_1-k^{\prime}_2)$ \\  \hline   
  $t_{1s}  = L(k^{\prime}_1+k^{\prime}_s)$     &  $\tau_{1s} = L(k^{\prime}_1-k^{\prime}_s)$   \\ \hline   
  $t_{1i}  = L(k^{\prime}_1+k^{\prime}_i)$       &$\tau_{1i}  = L(k^{\prime}_1-k^{\prime}_i)$  \\  \hline 
   $t_{2s} = L(k^{\prime}_2+k^{\prime}_s) $       & $ \tau_{2s}  = L(k^{\prime}_2-k^{\prime}_s)$ \\  \hline 
      $t_{2i} = L(k^{\prime}_2+k^{\prime}_i) $       & $ \tau_{2i}  = L(k^{\prime}_2-k^{\prime}_i)$ \\  \hline 
  \hline  
\end{tabular}\quad
\begin{tabular}{| c | c | }
  \hline 
  $T_s=\frac{\sigma_2^2}{\sigma_1^2+\sigma_2^2}t_{12}$     & $T_{i}=-\frac{\sigma_1^2}{\sigma_1^2+\sigma_2^2}t_{12}$   \\  \hline
  $t_{12}  = L(k^{\prime}_1+k^{\prime}_2)$    &  $\tau_{12}= L(k^{\prime}_1-k^{\prime}_2)$ \\  \hline   
  $t_{1s}  = 2 L k^{\prime}_1 $     &  $\tau_{1s} =0$   \\ \hline   
  $t_{1i}  = t_{12}$       &$\tau_{1i}  = \tau_{12}$  \\  \hline 
   $t_{2s} =t_{12} $       & $ \tau_{2s}  =-\tau_{12}$ \\  \hline 
      $t_{2i} = 2 L k^{\prime}_2 $       & $ \tau_{2i}  = 0$ \\  \hline 
  \hline  
\end{tabular}
\caption{Temporal parameters in analytical expressions for two cases.   Left:  general case, for which the four waves could involve different polarizations and propagation modes, right: identical polarizations and propagation modes for all four waves; note that the definition $k^{\prime}_{\mu}\equiv k^{\prime}_{\mu}(\omega_{\mu}^0)$ is used throughout.}\label{Tpar}
\end{center}
\end{table}}

Note that for a sufficiently large time of arrival difference between the two pump pulses at the opposite fiber ends $\tau$,  leading to the condition $\Lambda \gg 1$, the pump pulses overlap temporally outside the fiber and the process ceases to occur.  If this time of arrival difference $\tau$ vanishes, the value of $\Lambda$ tends to be small since it is given by the ratio of the transit time \emph{difference} through the fiber of the two pumps, divided by  transit time \emph{sum};  thus, often we may approximate $\Lambda \approx 0$.

\begin{figure}[t]
\begin{center}
\centering
\includegraphics[width=14cm]{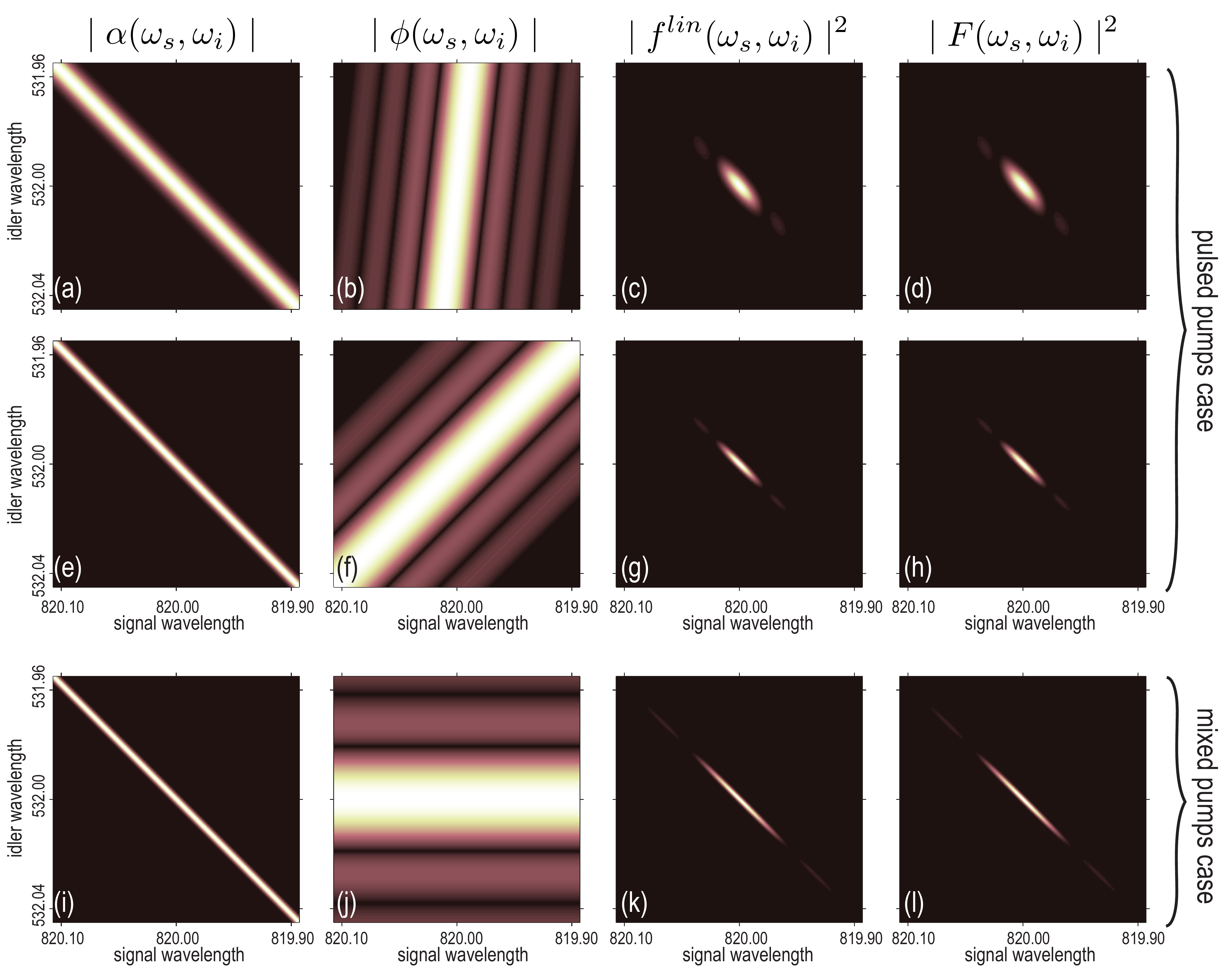}
\caption{Spectral correlation properties of CP-SFWM two-photon states, assuming a step index fiber (core radius $r=1.5\mu$m and $NA=0.13$) with $L=1$cm. (a)-(d) Pulsed pumps case with $\sigma_1=0.01THz$ and $\sigma_2=0.03THz$.  (e)-(h) Pulsed pump case with $\sigma_1=\sigma_2=0.01THz$. (i)-(l) Mixed pump case with $\sigma=0.01THz$. $\alpha(\omega_s,\omega_i)$ is the pump envelope function (given by equation (\ref{alpha_Pul}) for the pulsed case and by equation (\ref{pump_analt_M}) for the mixed case). $\phi(\omega_s,\omega_i)$ is the phasematching function (given by equation (\ref{phi_Pul}) for the pulsed case and by equation (\ref{PM_analt_M}) for the mixed case). $f^{lin}(\omega_s,\omega_i)$ represents the JSI under the linear $\Delta k$ approximation. $F(\omega_s,\omega_i)$ represents the JSI without resorting to approximations  (calculated numerically from equation (\ref{JSAp}) for the pulsed case and from equation (\ref{JSAm}) for the mixed case).} 
\label{JSIsynthesis}
\end{center}
\end{figure}

Note that the assumptions (see first paragraph of this section) used for the derivation of the approximate expression for the joint spectral amplitude $f^{lin}_{\it{P}}(\nu_s,\nu_i)$  are 
no longer valid for sufficiently large signal and idler spectral spreads around the central SFWM frequencies $\omega_s^0$ and $\omega_i^0$.  It is worth pointing out that for the specific source designs presented in this paper (see Figs. \ref{JSIsynthesis} and \ref{JSIfactSyn}, below) these approximations are well justified:  plots of the joint spectrum derived from the expression $|f^{lin}_{\it{P}}(\nu_s,\nu_i)|^2$ are in excellent agreement with plots derived from direct numerical integration, without resorting to approximations,  according to Eq. (\ref{JSAp}).  

The same assumptions considered in the derivation of $ f^{lin}_{\it{P}}(\nu_s,\nu_i)$ can be applied in equation (\ref{NumPhot_P}) in order to get a closed analytical expression of the emission rate, which leads to 

\begin{equation}
\label{N_analt_P} N_{\it{P}}^{lin}=\frac{2^5n_{1}n_{2}c^2\gamma^2p_1p_2h(\omega_s^0,\omega_i^0)}{R(k^{\prime}_1+k^{\prime}_2)( k^{\prime}_s+k^{\prime}_i)\omega_{1}^0\omega_{2}^0}\left[\mbox{erf}\left(\frac{1+\Lambda}{2\sqrt{2}B}\right) +\mbox{erf}\left(\frac{1-\Lambda}{2\sqrt{2}B}\right)\right].
\end{equation}

From the above equation, using the property that the $\mbox{erf}(x)$ function saturates to a value of $1$ for $x \gtrsim 2$ (or to a value of $-1$ for $x \lesssim -2$), we may show that there exists an effective fiber length $L_{eff}$ given by 

\begin{equation}
\label{Leff} L_{eff}=\frac{4\sqrt{2}\sqrt{\sigma_1^2+\sigma_2^2}}{(1+\Lambda)(k^{\prime}_1+k^{\prime}_2)\sigma_1\sigma_2}
= \frac{4\sqrt{2}\sqrt{\Delta t_1^2+\Delta t_2^2}}{(1+\Lambda)(k^{\prime}_1+k^{\prime}_2)},
\end{equation}

\noindent  where $\Delta t_1\equiv 1/ \sigma_1$ and $\Delta t_2\equiv 1/ \sigma_2$ represent the temporal durations for pump $1$ and pump $2$, respectively,  with the property that increasing the fiber length beyond $L=L_{eff}$ does not lead to any further increase of the source brightness; thus, $L_{eff}$ corresponds to the maximum interaction length.   Physically, $L_{eff}$ represents the length of fiber over which the two pumps overlap temporally.  Note that making one of the two pumps approach the continuous wave (monochromatic) limit implies that the interaction length can increase without limit, which as is described below is helpful for the optimization of the source brightness. 

Let us consider a case where both pumps have a non-zero bandwidth; we can then write $B=\sqrt{1+r}/(\sigma_1 t_{12})$, or  $B = \sqrt{1+r}\Delta t_1 / t_{12}$, (with $r \equiv \sigma_1^2/\sigma_2^2$).   Thus, if $\Delta t_1$  is much smaller than the sum of the transit times through the fiber of the two pump pulses, represented by $t_{12}$,  then $B$ can be a  small number.  Essentially, a small $B$ implies that since the interaction length is much shorter than the fiber length, the two-photon state is free from any effects related to the air-fiber and fiber-air interfaces.   In this $B\to0$ limit, which can always be reached through a combination of pulsed pumps with a sufficiently small pump duration together with a sufficiently long fiber, the phasematching function becomes $\phi_{\it{P}}(x)\to 2\mbox{exp}(-B^2x^2)$.    This limit is interesting for applications where the suppression of the sinc-function sidelobes is beneficial, as is the case for the generation of very high-quality factorable states.

Let us now consider the case where the pump bandwidths are highly unbalanced.  In particular, the condition  $\sigma_1 \ll \sigma_2$ leads to $B \approx 1/(t_{12} \sigma_1)$, while  similarly $\sigma_2 \ll \sigma_1$ leads to $B \approx 1/(t_{12} \sigma_2)$.   In the limit where the smaller of the two bandwidths becomes very small, the value of $B$ becomes very large, in which case it may be shown that the phasematching function becomes $\phi_{\it{P}}(x)\to\mbox{sinc}(x/2)$.  In practice, for values $B \gtrsim 1.0$, the phasematching function is already well described by a sinc function; in this regime, unlike for the small $B$ limit,  the fiber edges play an essential role.

\begin{figure}[t]
\begin{center}
\centering
\includegraphics[width=14cm]{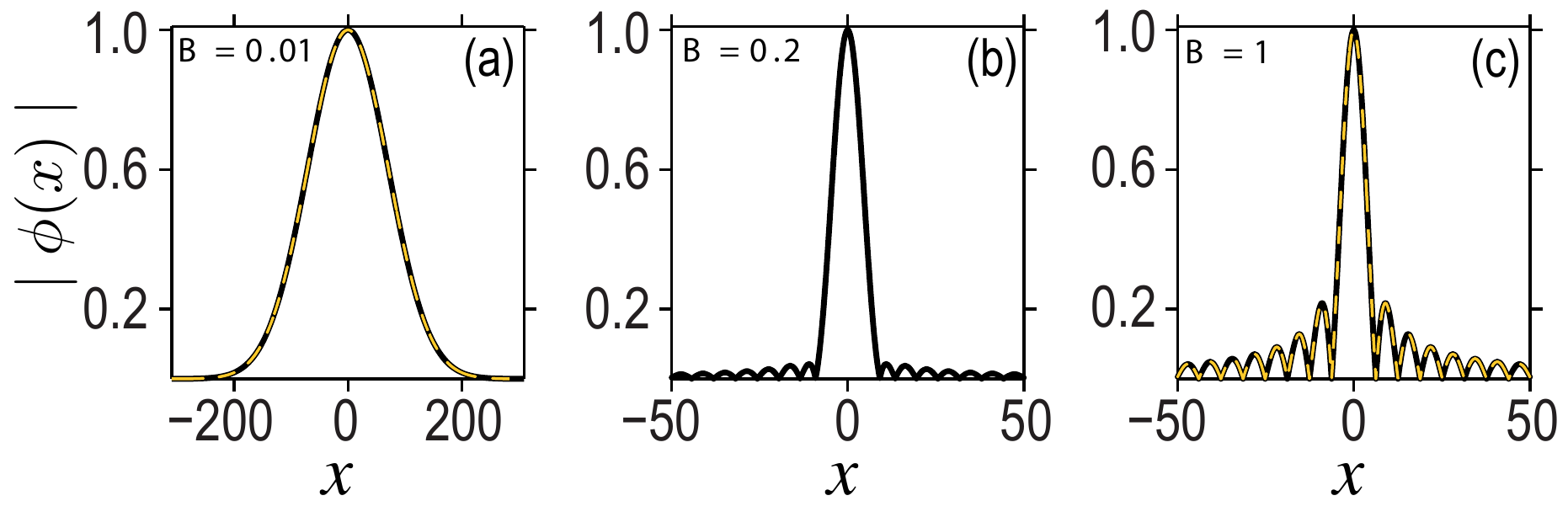}
\caption{Absolute value of the phasematching function [see equation  (\ref{phi_Pul})] for different values of parameter $B$, with $\Lambda=-0.00685$ (corresponding to pumps at 532nm and 820nm with $\tau=0$, assuming a step-index fiber with core radius $r=1.5 \mu$m and numerical aperture $NA=0.13$).}
\label{PMfunc}
\end{center}
\end{figure}

The behavior  of the function $ |\phi_{\it{P}}(x)|$ as parameter $B$ varies is summarized in Fig. \ref{PMfunc}.   On the one hand, panel (a) illustrates the small $B$ limit (in this case with $B=0.01$), in which the function  $|\phi_{\it{P}}(x)|$ becomes the Gaussian function  $\mbox{exp}(-B^2x^2)$.   On the other hand, panel (c) illustrates the large $B$ limit (in this case with $B=1$) in which the function  $|\phi_{\it{P}}(x)|$ becomes $\mbox{sinc}(x/2)$.   In both of these panels, the $|\phi_{\it{P}}(x)|$ function as given by Eq. (\ref{phi_Pul}) is plotted with a solid black line, while the Gaussian or sinc limiting behaviors are plotted with a dashed yellow line.  It becomes evident that there is an excellent agreement between these.  In panel (b) we show an intermediate case with $B=0.2$ for which  $|\phi_{\it{P}}(x)|$ is not well described neither by a Gaussian nor by a sinc function.

Let us now analyze how the functions $\alpha_P(\nu_s,\nu_i)$ and $\phi_P(\nu_s,\nu_i)$ define the joint spectral intensity of CP-SFWM photon pairs.  It is clear from equations (\ref{alpha_Pul}) and (\ref{phi_Pul}) that while $\alpha_P(\nu_s,\nu_i)$ is oriented at $-45^{\circ}$ in $\{\omega_s,\omega_i\}$ space with a width $\sqrt{\sigma_1^2+\sigma_2^2}$, the orientation and width of $\phi_{\it{P}}(\nu_s,\nu_i)$ depend, both, on pump and fiber parameters. The orientation angle of the function $\phi_{\it{P}}(\nu_s,\nu_i)$ is given by

\begin{equation}
\label{ang}
\theta_{si}=\arctan\left(-\frac{T_s}{T_i}\right)=\arctan \left(\frac{\sigma_2^2}{\sigma_1^2}\right),
\end{equation}

\noindent with the last equality valid for the case where all four waves have the same polarization and transverse mode.  Note that $\sigma_2^2/\sigma_1^2 \ge 0$, so that $\theta_{si}$ is constrained as $0\le\theta_{si}\le90^\circ$, i.e. the function $\phi_{\it{P}}(\nu_s,\nu_i)$ has contour curves with non-negative slope, including the two limiting cases of horizontal and vertical orientations.  As regards the width of the function $\phi_{\it{P}}(\nu_s,\nu_i)$, it is helpful to consider separately the limiting cases for large $B$ where this function is well described by $\mbox{sinc}(x/2)$, and for small $B$ for which this function becomes $\exp(-B^2 x^2)$.   In the first case, the width is inversely proportional to the fiber length $L$ (since both $T_s$ and $T_i$ in $x=T_s \nu_s+T_i \nu_i$ are linear in $L$).  In the second case, the width no longer depends on $L$ (since $B$ is proportional to $L^{-1}$ while $x$ is proportional to $L$).  Thus, as the fiber length is increased the width of the function $\phi_{\it{P}}(\nu_s,\nu_i)$ diminishes, eventually the shape turning Gaussian at which point the width and shape of the function $\phi_{\it{P}}(\nu_s,\nu_i)$ no longer responds to further increasing $L$.  Thus, increasing $L$ beyond the length defined by non-zero temporal overlap between the two pump pulses, $L_{eff}$, has no effect \emph{neither} on the flux \emph{nor} on the joint spectral intensity.

The discussion of the previous paragraph is illustrated in figure \ref{JSIsynthesis}, in which we show the two-photon state obtained for  a step-index fiber (with core radius $r=1.5 \mu$m and numerical aperture $NA=0.13$) with length $L=1$cm and two different pulsed pumps configurations: i) $\sigma_1=0.01THz$ and $\sigma_2=0.03THz$, panels (a)-(d), for which $B=1.07$; ii) $\sigma_1=\sigma_2=0.01$THz, panels (e)-(h), for which $B=1.43$.  In this block of figures, the function $\alpha_P(\nu_s,\nu_i)$ is shown in the first column, the function $\phi_{\it{P}}(\nu_s,\nu_i)$ in the second column, the JSI $|\alpha_P(\nu_s,\nu_i)\phi_{\it{P}}(\nu_s,\nu_i)|^2$ in the third column, while the numerically-calculated JSI is shown in the fourth column.  Note that while the third  column corresponds to the analytical joint spectral intensity, defined as $|f^{lin}_{\it{P}}(\nu_s,\nu_i)|^2$, the fourth column was obtained by numerical integration of equation (\ref{JSAp}) without resorting to the linear approximation of the $\Delta k$ function.  It is evident that the approximate analytical results agree extremely well with the numerically-calculated ones.   Also, consistent with Eq. (\ref{ang}), while the orientation of the function $\phi_{\it{P}}(\nu_s,\nu_i)$ is $45^{\circ}$ for equal pump bandwidths, it approaches a vertical orientation for unequal pump banwidths $\sigma_2>\sigma_1$, and becomes fully vertical for $\sigma_2 \gg \sigma_1$.  Similarly, (not shown in the figure), for $\sigma_2<\sigma_1$ the function $\phi_{\it{P}}(\nu_s,\nu_i)$ approaches a horizontal orientation while it becomes fully horizontal for $\sigma_2\ll\sigma_1$.

For highly unbalanced pump bandwidths, and in particular when one of the two pumps approaches the monochromatic limit, the interaction length between the two pump pulses increases, in principle, without limit.  Such a mixed pumps configuration could have important implications for the ability to reach high emission rates.  

Following a similar treatment as used above for the pulsed pumps case, it can be shown that the JSA function given in equation (\ref{JSAm}) can be expressed, under the linear $\Delta k$ approximation, as $ f^{lin}_{\it{M}}(\nu_s,\nu_i)=\alpha_{\it{M}}(\nu_s,\nu_i)\phi_{\it{M}}(\nu_s,\nu_i)$, with the pump envelope function $\alpha_{\it{M}}(\nu_s,\nu_i)$ and the phasematching function $\phi_{\it{M}}(\nu_s,\nu_i)$ given by

\begin{equation}
\label{pump_analt_M}  \alpha_{\it{M}}(\nu_s,\nu_i)=\mbox{exp}\left[-\frac{(\nu_s+\nu_i)^2}{\sigma^2}\right],
\end{equation}

\begin{equation}
\label{PM_analt_M}  \phi_{\it{M}}(\nu_s,\nu_i)=\mbox{sinc}\left[\frac{1}{2}(\tau_{1s}\nu_s+t_{1i}\nu_i)\right]\mbox{exp}\left[i t_{1s}\nu_s+i t_{1i}\nu_i\right],
\end{equation}

\noindent where $\sigma$ is the bandwidth of the pulsed pump and $\tau_{1s}$, $t_{1s}$, and $t_{1i}$ are defined in table \ref{Tpar}. Likewise, it can be demonstrated by integration of equation (\ref{NumPhot_M}), and under the linear phasemismatch approximation, that the emitted flux can be expressed as

\begin{equation}
\label{N_analt_M} N_{\it{M}}^{lin}=\frac{2^6n_{1}n_{2}c^2\gamma^2p_1p_2Lh(\omega_s^0,\omega_i^0)}{\omega_{1}^0\omega_{2}^0| k^{\prime}_s+k^{\prime}_i|},
\end{equation}

\noindent where the dependence on the various experimental parameters of the emission rate appears explicitly. Particularly, it can be seen, as expected, that $N_{\it{M}}$ increases linearly with the fiber length, indicating that the interaction length is not capped as it is for the pulsed pumps configuration. 

\begin{figure}[t]
\begin{center}
\centering
\includegraphics[width=16cm]{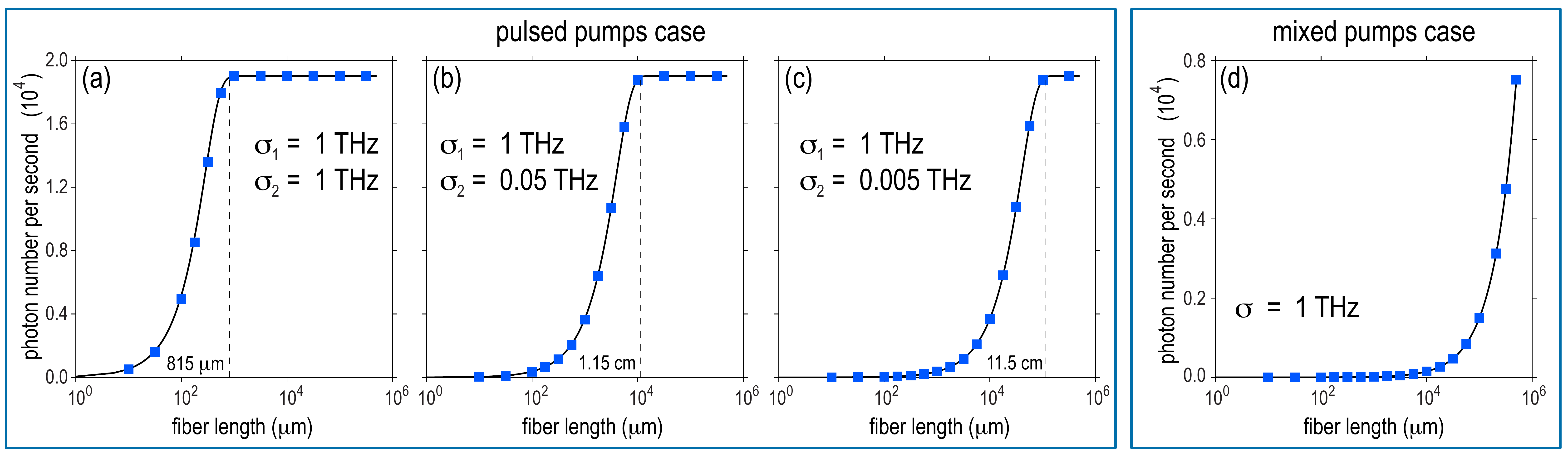}
\caption{Emitted flux as a function of the fibre length. (a)-(c) Pulsed pumps configurations. Vertical dashed lines indicate the effective length $L_{eff}$ (with the  $L_{eff}$ values shown).  (d) Mixed pumps configuration. The blue markers were obtained from numerical evaluation of equations (\ref{NumPhot_P}) and (\ref{NumPhot_M}) for the pulsed and mixed pumps configuration, respectively, while the black solid lines are the analytical results obtained under the linear phasematching approximation, according to Eqns. (\ref{N_analt_P}) and (\ref{N_analt_M}).} 
\label{brightness}
\end{center}
\end{figure}

In figures \ref{JSIsynthesis}(i) to \ref{JSIsynthesis}(l) we illustrate the two-photon state obtained for a CP-SFWM source in the mixed pumps configuration, for which we have assumed the same parameters as in figures \ref{JSIsynthesis}(e) to \ref{JSIsynthesis}(h), except that pump $2$ is now monochromatic ($\sigma_2\to 0$); in this case the contours of the phasematching function become horizontal.  Note that for the mixed pumps configuration  the dependence of the spectral envelope  function and the phasematching function on the parameters of the source become decoupled; i.e., the width of $ \alpha_{\it{M}}(\nu_s,\nu_i)$ is proportional to the pulsed pump bandwidth  (with an orientation at $-45^\circ$), while the width of $\phi_{\it{M}}(\nu_s,\nu_i)$ depends on fiber properties, including its length $L$ and dispersion (with a horizontal orientation) . 

In Figure \ref{brightness} we illustrate the behaviour of the source brightness as a function of the fiber length, for both the pulsed and mixed pumps cases.    In panels (a)-(c) we show this behaviour for three different values of $\sigma_2$ (1 THz, 0.05 THz, and 0.005 THz, respectively), and for a fixed value $\sigma_1=1$ THz, where a vertical dashed line indicates the effective length $L_{eff}$;  it may be appreciated that the brightness reaches a plateau at $L \approx L_{eff}$. In panel (d) we show the corresponding behaviour for the mixed pumps scheme, for $\sigma=1$THz; note that in this case the brightness grows linearly with $L$ without saturating to a fixed value.

\section{\label{fact}Factorable two-photon states generation}

In this section we will show that when restricting our discussion to a co-polarized CP-SFWM process implemented in a fiber which supports a single transverse mode, a factorable state can be obtained for \emph{any} phasematched configuration, with frequencies such that $\omega_1=\omega_s$ and $\omega_2=\omega_i$.   

In a SFWM process for which all four waves propagate in the same polarization/transverse spatial mode, quantum entanglement can  reside only in the spectral degree of freedom. Spectral correlation properties are then governed by the joint spectrum of the two-photon state, see equations (\ref{JSAp}) and (\ref{JSAm}).  In order to facilitate the analysis we focus here on the analytical expressions of the JSA based on the linear approximation of $\Delta k$, which were introduced in section \ref{sec.analytical}, $f^{lin}_{\it{P}}(\nu_s,\nu_i)$ and $ f^{lin}_{\it{M}}(\nu_s,\nu_i)$ for the pulsed and mixed pumps configurations, respectively. In both cases, as discussed above, spectral correlations are determined by the relative orientation and spectral widths of the pump envelope and phasematching functions, see equations (\ref{alpha_Pul}) and (\ref{phi_Pul}) for the pulsed case, and equations (\ref{pump_analt_M}) and (\ref{PM_analt_M}) for the mixed case. The two-photon states becomes factorable if the JSA function is separable, i.e. if it can be written as $f(\omega_s,\omega_i)=S_s(\omega_s)I_i(\omega_i)$. 

Let us consider the limit $B\rightarrow 0$, which as discussed in section \ref{sec.analytical} can always be attained for a combination of sufficiently short pump pulses and for a sufficiently long fiber.  
In this case, the joint spectral intensity  $I(\nu_s,\nu_i) \equiv |\alpha_P(\nu_s,\nu_i)\phi_P(\nu_s,\nu_i)|^2$ may be expressed as 

\begin{eqnarray} 
\label{JSAfactpuls}
I(\nu_s,\nu_i) &\propto \exp[-2B^2(T_s \nu_s+ T_i \nu_i)^2]\exp\left[-\frac{2(\nu_s+\nu_i)^2}{\sigma_1^2+\sigma_2^2}\right] \nonumber \\
&= \exp \left[ -\frac{2\nu_s^2}{\sigma_1^2} \right] \exp \left[ -\frac{2\nu_i^2}{\sigma_2^2} \right].
\end{eqnarray}

Note that in order to write down the last equality, we have used the expressions for $B$, $T_s$, and $T_i$ valid for the case where all four waves propagate in the same polarization/transverse spatial mode (see table \ref{Tpar}). This result, valid in the limit $B \rightarrow 0$, is remarkable on a number of fronts: i) the two-photon state is \emph{automatically factorable}, in addition to the underlying phasematching condition being attained automatically, as already discussed in section \ref{sec.PM},  ii) the state becomes completely independent of fiber parameters and only depends on the two pumps, and iii) the bandwidth of the signal photon is identical to the pump $1$ bandwidth, while the bandwidth of the idler photon is identical to the pump $2$ bandwidth.  

By direct plotting of the  $\phi_P(x)$ function, we may verify that for $B \lesssim 0.14$ this function is essentially identical to $\exp(-B^2 x^2)$; this corresponds to the regime under which Eq. (\ref{JSAfactpuls}) is valid.  This leads to the following factorability condition, 

\begin{equation} 
\label{Lumbral}
L \gtrsim \frac{(0.14)^{-1}\sqrt{\sigma_1^2+\sigma_2^2}}{(k_1'+k_2')\sigma_1\sigma_2}=\frac{(0.14)^{-1}\sqrt{\Delta t_1^2+\Delta t_2^2}}{k_1'+k_2'}.
\end{equation}

Eq.~(\ref{Lumbral}) provides a threshold fiber length (which decreases as the pump temporal durations are reduced) so that if the fiber length exceeds this threshold the two-photon state is always factorable.  It is interesting to compare the factorability fiber length threshold (see eq.~(\ref{Lumbral})) with the maximum interaction length $L_{eff}$ (see Eq. (\ref{Leff})). Note that these two expressions are essentially identical; indeed, if the fiber becomes longer than the distance over which the two pump pulses are temporally overlapped, two effects are observed: i) the brightness can no longer increase, and ii) edge effects related to the air-fused silica interfaces disappear.  Thus, as $L$ is increased, the brightness plateaus at $L=L_{eff}$ and the state reaches the Gaussian factorable form as described by Eq. (\ref{JSAfactpuls}). Note that for ps pumps the values of $L_{eff}$ tend to be in the range of mm to cm making this scheme for factorable photon-pair generation highly practical.   
 
 \begin{figure}[t]
\centering
\includegraphics[width=15cm]{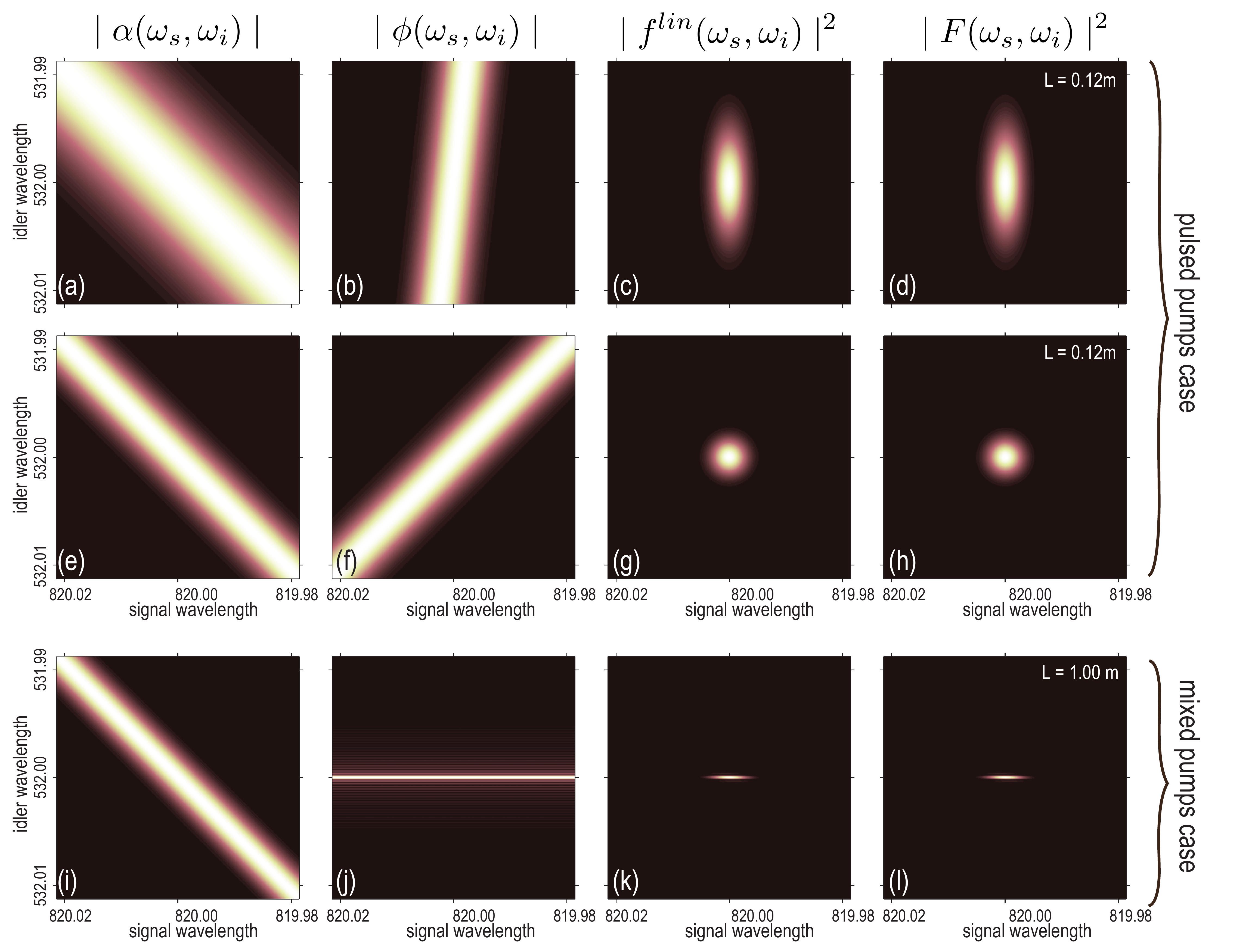}
\caption{Synthesis of the joint spectral intensity (obtained with the same fiber as assumed in Fig. 3)  for: (a)-(d) Pulsed pumps case with $\sigma_1=0.01THz$ and $\sigma_2=0.03THz$, $L=0.12$m.  (e)-(h) Pulsed pump case with $\sigma_1=\sigma_2=0.01THz$, $L=0.12$m. (i)-(l) Mixed pump case with $\sigma=0.01THz$, and $L=1$m. The specific fiber lengths considered here are longer than the threshold lengths in Eq. (\ref{Lumbral}) and  Eq. (\ref{LumbralCW}) for the pulsed and mixed pumps configurations, respectively.}
\label{JSIfactSyn}
\end{figure}
 
As one or both of the pump bandwidths are reduced, the effective length $L_{eff}$ increases without limit. Thus, in the limit where either $\sigma_1 \rightarrow 0$ and/or $\sigma_2 \rightarrow 0$, the interaction length can become arbitrarily large, in practice limited by the fiber length, and Eqns. (\ref{JSAfactpuls}) and (\ref{Lumbral}) derived above for the pulsed pumps case can no longer be applied. Thus, let us now consider the question of factorability in the mixed pumps case, for which pump $2$ is monochromatic, and pump $1$ has a certain non-zero bandwidth $\sigma$.  In this case, the joint spectral intensity $I(\nu_s,\nu_i) \equiv |\alpha_M(\nu_s,\nu_i)\phi_M(\nu_s,\nu_i)|^2$ may be expressed as

\begin{eqnarray} 
\label{JSIcwana}
I(\nu_s,\nu_i) &\propto \left(\mbox{sinc}\left[ \frac{1}{2}(\tau_{1s} \nu_s+ t_{1i} \nu_i )\right]\right)^2
\exp\left[-\frac{2(\nu_s+\nu_i)^2}{\sigma^2}\right] \nonumber \\
&=  \left(\mbox{sinc}\left[ \frac{t_{12} \nu_i }{2}\right]\right)^2
\exp\left[-\frac{2(\nu_s+\nu_i)^2}{\sigma^2}\right].
\end{eqnarray}

Note that in order to write down the last equality, we have used the expressions for $\tau_{1s}$ and $t_{1i}$ valid for the case where all four waves propagate in the same polarization / transverse spatial mode (see table \ref{Tpar}).  This JSI $I(\nu_s,\nu_i)$ then becomes factorable if the width of the sinc function, along $\nu_i$, is much less than the width of the exponential function, along $\nu_s+\nu_i$.   With the help of the Gaussian approximation $\mbox{sinc}(x)\approx \exp(-\Gamma x^2)$ (with $\Gamma=0.193$), we then arrive at the following condition for factorability

\begin{equation}
\label{LumbralCW}
L \gg \frac {2 \Delta t}{\sqrt{\Gamma}(k_1'+k_2')},
\end{equation}

\noindent which makes it clear that for a sufficiently long fiber, the two-photon state becomes factorable; note that in this equation $\Delta t \equiv \sigma^{-1}$.  Note that because the sinc function depends only on the frequency $\nu_i$, the sidelobes associated with this function  will run parallel to the $\nu_i$ axis and will not, therefore, introduce correlations (this observation also serves to justify the use of the Gaussian approximation).    In this limit, we may set $\nu_i \rightarrow 0$ in the exponential term, so that the joint spectral intensity can be approximated as

\begin{eqnarray} 
\label{jsiCWfact}
I(\nu_s,\nu_i) 
\approx \left(\mbox{sinc}\left[ \frac{t_{12} \nu_i }{2}\right] \right)^2
\exp\left[-\frac{2\nu_s^2}{\sigma^2}\right].
\end{eqnarray}

It is remarkable that for, both, the pulsed pumps and the mixed pumps configurations a factorable state can always be reached for a sufficiently long fiber. This behaviour is illustrated in figure \ref{JSIfactSyn} in which, for the same pump configurations as in figure \ref{JSIsynthesis}, we show the synthesis of the joint spectral intensity for fiber lengths longer than the threshold lengths in Eq. (\ref{Lumbral}) and  Eq. (\ref{LumbralCW}) for the pulsed and mixed pumps configurations, respectively. It is evident in this figure that the three source scenarios lead to factorable two-photon states. It is worth emphasizing that while in the case of standard (co-propagating) SFWM, factorability demands specific combinations of fiber length and pump bandwidth \cite{Garay07}, for CP-SFWM the factorability conditions are considerably more relaxed and in fact \emph{all} phasematched configurations can lead to a factorable state for a sufficient fiber length.  

\begin{figure}[t]
\begin{center}
\centering
\includegraphics[width=13cm]{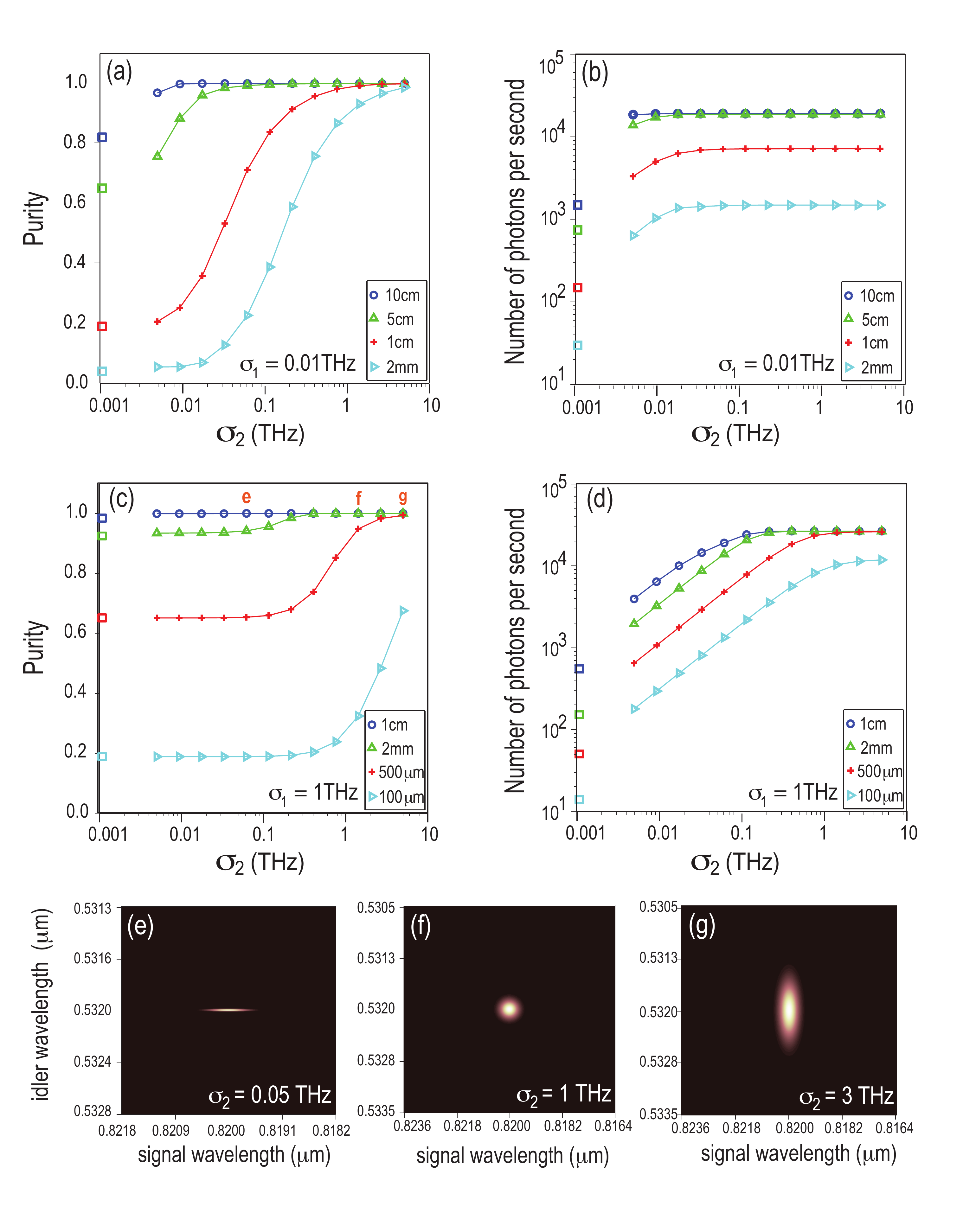}
\caption{(a) Purity versus $\sigma_2$  as a function of fiber length with  $\sigma_1=0.01THz$. (b) Photon-pair emission rate vs $\sigma_2$ as a function of fiber length. The average power of the two pumps is 50mW. (c) and (d) Similar to (a) and (b), but with $\sigma_1=1THz$. (e)-(g) Joint spectral intensity for the $\sigma_2$ values indicated as e, f, g on panel (c) and $L=1$cm. For all three cases, the numerically evaluated purity reaches very close to unity.}
\label{PurityVsSigma2}
\end{center}
\end{figure}

In order to quantify the degree of factorability of CP-SFWM photon pairs, we evaluate the heralded-single-photon state purity $p \equiv\mbox{Tr}(\hat{\rho}_s^2)=1/K$ in terms of the Schmidt number $K$, where $\hat{\rho}_s$ is the reduced density operator for the signal state \cite{URen05}. Thus, an ideal factorable two-photon state is related to an ideal single-photon purity $\mbox{Tr}(\hat{\rho}_s^2)=1$. In figure \ref{PurityVsSigma2}(a) we show the numerically-calculated purity as a function of $\sigma_2$, while $\sigma_1$ and $L$ remain fixed; results are shown for four different fiber lengths, as indicated, and $\lambda_{1}=0.820\mu$m, $\lambda_{2}=0.532\mu$m, and $\sigma_1=0.01THz$.  Square markers in the figure correspond to the purity obtained for the mixed pump case, for which $\sigma_2\to0$, see equation (\ref{JSAm}).  Figure \ref{PurityVsSigma2}(b) shows the number of photon pairs emitted per second for the same parameters assumed in panel (a).  Panels (c) and (d) are similar to (a) and (b), except for a larger value of the pump 1 bandwidth: $\sigma_1=1$THz.  From these plots the following two behaviors as the fiber length is increased become apparent: i) the two-photon state becomes increasingly factorable, and ii) the source brightness reaches a plateau.   In addition, increasing $\sigma_2$ leads to a reduced effective length $L_{eff}$, thus boosting the purity, for a given value of $L$.  Note from Fig. \ref{PurityVsSigma2} (b) and (d)  that while the use of very short fibers would lead to the need for large pump bandwidths in order to attain factorability, with correspondingly larger self-/cross-phase modulation effects,  there is no need in practice to use such short, e.g. sub-mm, fibers which are in addition comparatively more challenging to handle. 

In Fig.\ref{PurityVsSigma2}(c) we have indicated with the letters e, f, and g three particular choices of parameters, which lead to the joint spectra shown in Fig.\ref{PurityVsSigma2}(e), (f), and (g).  Note that for all of these three parameter choices, the two-photon state is essentially factorable.    

As a final remark in this section, it is worth  mentioning that in the intermodal CP-SFWM configuration, discussed at the end of section \ref{sec.PM}, the factorability of the two-photon state is preserved as compared with the case in which all interacting fields propagate in the fundamental mode, regardless of the higher-order fiber mode employed. The emission rate, however, may be compromised as the order of the excited mode used increases, due to a reduced overlap between the interacting modes.

\section{\label{fact}Ultra-narrowband single-photon wavepacket generation}

Atom-photon interfaces rely on the ability of a single photon to be absorbed by a single atom; such interfaces involve matching both frequency and bandwidth of the single photons to the intended atomic transition in a given atomic species. While such electronic transitions typically have bandwidths in the region of MHz, the natural bandwidths of SPDC and SFWM sources tend to be many orders of magnitude greater. A possible solution is to place the nonlinear medium responsible for photon-pair generation inside a high-finesse cavity so as to restrict the emission bandwidth as needed, without adversely affecting the source brightness \cite{Garay13}.

Let us observe from Eq. (\ref{JSAfactpuls}) that in the pulsed pumps configuration, specifically in the regime $L>L_{eff}$ for which the JSI becomes factorable and fully Gaussian,  the emission bandwidths are `inherited' from the pumps: i.e $\sigma_s=\sigma_1$ and $\sigma_i=\sigma_2$. This is a reflection of the achromatic phasematching for which the fiber dispersion experienced by the signal and pump $1$, on the one hand, and by the idler and pump $2$, on the other hand, cancel each other out so that the two-photon state is determined exclusively by the pumps.   As one or both of the pump bandwidths approach the monochromatic limit, the effective length $L_{eff}$ becomes infinite and the expression in Eq. (\ref{JSAfactpuls}) for the two photon state is no longer valid.    

Let us then consider the possibility of generating photon pairs in the mixed pumps configuration, for which at least one of the two pumps exhibits a very narrow bandwidth.   For a sufficient fiber length (obeying Eq. (\ref{LumbralCW})), the joint spectral intensity is given by Eq. (\ref{jsiCWfact}).  Let us observe that in this regime, the bandwidth of the signal photon `inherits' the bandwidth of pump $1$, i.e. $\sigma_s=\sigma$,  as occurs for the pulsed pumps case.  However, note that the bandwidth of the idler photon $\sigma_i$ is determined not by the pumps but solely by fiber properties, and with the help of the Gaussian approximation, can be expressed as

\begin{equation}
\label{bandwI}
\sigma_i=\frac{2}{\sqrt{\Gamma} L (k_1'+k_2')}.
\end{equation}

It is important to point out a key difference with respect to standard co-propagating SFWM.  While for standard SFWM, the spectral properties are determined by reciprocal group velocity \emph{difference} coefficients of the form $L(k_p'-k')$, for CP-SFWM the spectral properties are replaced by reciprocal group velocity \emph{sum} coefficients  of the form $L(k_1'+k_i')=L(k_1'+k_2')$.   These reciprocal group velocity sum coefficients correspond to the sum of transit times for the  pump 1 and pump 2 waves through the fiber, as opposed to transit time differences as appear in the case of standard SFWM. The fact that the sum coefficients tend to be orders of magnitude greater than the difference counterparts has a profound implication:  because the idler bandwidth is inversely proportional to this reciprocal group velocity sum (difference) coefficient for counter-propagating (standard) SFWM, the resulting bandwidths are orders of magnitude smaller than for a comparable standard co-propagating source, as a direct consequence of the counter-propagating geometry. In practice this leads to the possibility of obtaining extremely small idler bandwidths for reasonable lengths of fiber. 

Suppose that a bandwidth $\delta\omega$ is desired for the idler photon.  We can then show from Eq. (\ref{bandwI}) that the fiber length which guarantees such a bandwidth is given by

\begin{equation}
L=\frac{2}{\sqrt{\Gamma} \delta\omega}\frac{1}{ k_1'+k_2'}.
\end{equation}

\begin{figure}[t]
\begin{center}
\centering
\includegraphics[width=12cm]{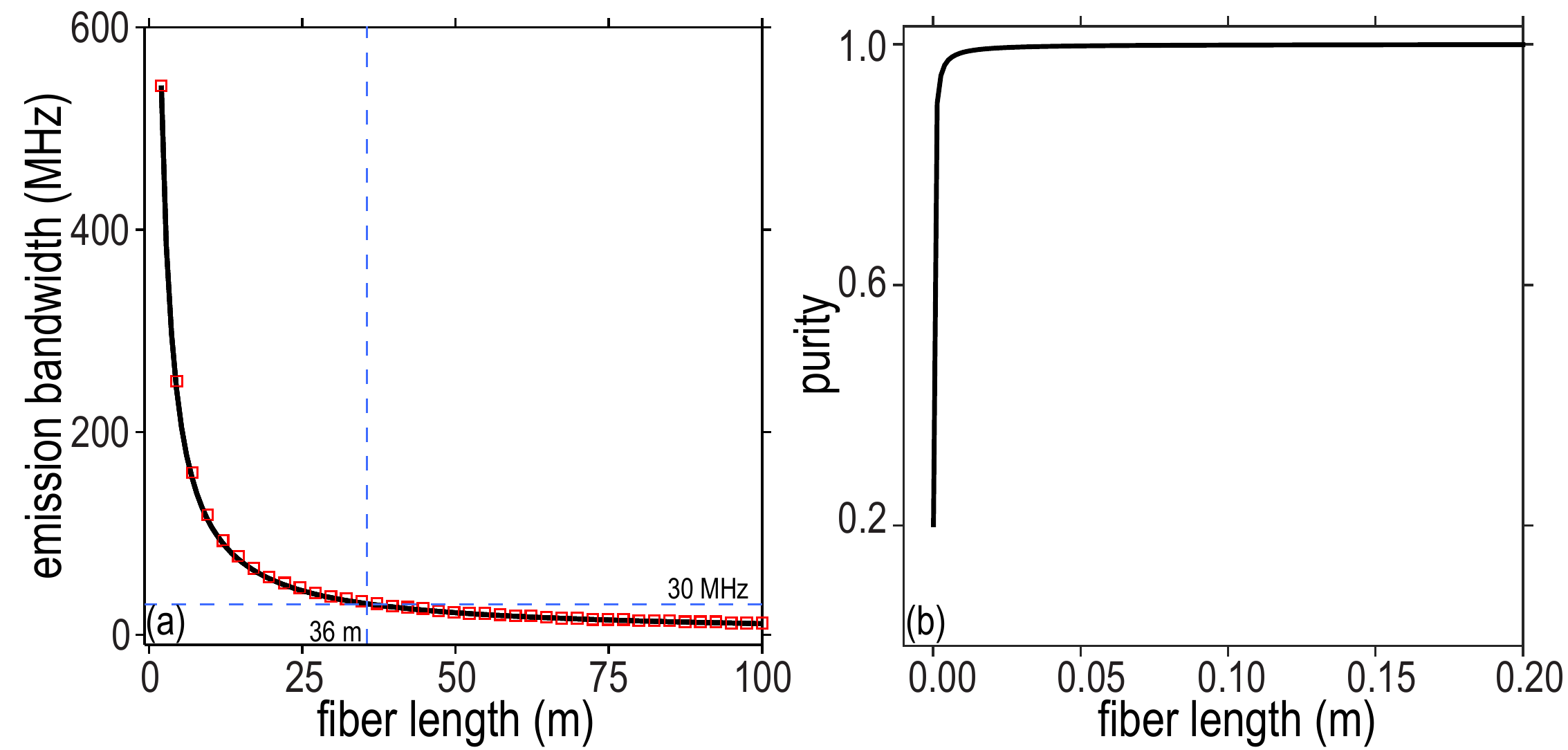}
\caption{Idler emission bandwidth (FWHM in intensity) (a) and purity (b) as function of fiber length $L$, obtained from CP-SFWM in the mixed pumps configuration. Results were evaluated assuming $\sigma=1$THz. Note from panel (b) that the two-photon state becomes essentially factorable for fiber lengths greater than the threshold length given in Eq. \ref{LumbralCW}, which in this case is around $0.5$mm. }
\label{BandW}
\end{center}
\end{figure}

In figure \ref{BandW}(a) we show results of the emission bandwidth $\sigma_i$ vs fiber length for a CP-SFWM source based on the mixed pumps scheme, with a pulsed pump of  $0.42$nm bandwidth centered at $0.820\mu$m (compatible with a picosecond Ti:Sapphire laser), and a monochromatic pump at $0.532\mu$m. The red squares represent results obtained numerically from equation (\ref{JSAm}), while the black solid line corresponds to those obtained analytically from equation (\ref{bandwI}). As indicated in the figure, for fiber lengths longer than $\sim36$m, (idler) single-photon wavepackets with bandwidths narrower than $30$MHz can be generated. In contrast, the bandwidth of the signal photon essentially equals that of the pump, i.e. $1.18$THz (FWHM).   In panel (b) of this figure we have shown the corresponding purity of a single idler photon vs fiber length, when heralded by the detection of a signal photon. 

The source scheme described above is suitable for applications in which it suffices for only one of the two photons in each pair to be narrow-band. Note that if both pumps are monochromatic it becomes possible, for a sufficiently long fiber, to generate photon pairs characterized by ultra-narrowband signal \emph{and} idler modes  (this case has not been analyzed in detail in this paper).

As has been emphasised, an important feature  of CP-SFWM  is the resulting phasematching achromaticity. Thus, for a given fiber it becomes possible to tune the emission frequencies  as controlled by the pump frequencies (with $\omega_s=\omega_1$ and  $\omega_i=\omega_2$), while preserving the emission bandwidths. This is a significant advantage in designing two-photon state sources. For example, a source may be designed so that one of the emission modes corresponds to a specific atomic transition (in frequency and bandwidth), while the other is tuned to the telecommunications band \cite{Fekete13}. 

\section{Conclusions}

In this paper we have described theoretically a new kind of spontaneous four wave mixing process, in which the two pump waves counter-propagate in the $\chi^{(3)}$ nonlinear medium, and in which the generated signal and idler photons likewise counter-propagate; we have referred to this process as counter-propagating spontaneous four wave mixing, or CP-SFWM. We have shown that in  this process, phasematching is attained \emph{automatically} regardless of the specific dispersion characteristics, leading to a signal frequency which equals the frequency of the pump wave travelling in the opposite direction, and likewise for the idler photon and the second pump wave. We have discussed that while a number of experimental aspects can slightly offset each of the generation frequencies with respect to the frequency of the corresponding pump wave, to aid discrimination of the SFWM photons from the pumps,  the use of an intermodal CP-SFWM process seems to be the most practical alternative.

We have presented two versions of the CP-SFWM process: in the first, which we refer to as the pulsed pumps configuration, both pumps are assumed to be pulsed while in the second, which we refer to as the mixed pumps configuration, one pump is assumed pulsed and the remaining pump is assumed to be monochromatic.   We have shown that in both of these cases, for an arbitrary phasematched source design, the state can always reach factorability for a sufficiently long fiber (or waveguide).  Moreover, the threshold length for factorability tends to be in the range of a few mm to a few cm, making the resulting automatic phasematching \emph{and} automatic factorability highly practical.   We have also shown that in the mixed pumps configuration,  the idler photon, which is emitted in counter-propagation to the monochromatic pump wave, can be made compatible in bandwidth with electronic transitions in atoms.  The latter eliminates the need for optical cavities, and is a direct consequence of the counter-propagating geometry for which the emission bandwidths are governed by the transit time \emph{sums} through the non-linear medium, rather than transit time \emph{differences} as in the case of standard SFWM. 

We point out that of the three properties discussed above, i.e. automatic phasematching, automatic factorability, and ultra-narrow single-photon bandwidths, at least the first two are amenable to integrated optics implementations, since they involve modest threshold lengths.  We believe that this new type of spontaneous four wave mixing process, in which the waves involved counter-propagate in the non-linear medium, may prove useful in future implementations of fiber- or waveguide-based photon-pair sources with engineered spatio-temporal properties.

\section{Acknowledgments}
This work was supported by CONACYT, M\'exico, PAPIIT (UNAM) grant number IN105915, and AFOSR grant FA9550-16-1-0458.

\section{References}
\bibliographystyle{unsrt}
\bibliography{References}

\end{document}